# High-pressure effects on the optical-absorption edge of $CdIn_2S_4$, $MgIn_2S_4$, and $MnIn_2S_4$ thiospinels

J. Ruiz-Fuertes[1], D. Errandonea[1,*], F. J. Manjón[2], D. Martínez-García[1],

A. Segura[1], V. V. Ursaki[3], and I. M. Tiginyanu[3]

[1]Dpto. de Física Aplicada-ICMUV, Universitat de València, Edificio de Investigación, c/Dr. Moliner 50, 46100 Burjassot (Valencia), Spain

[2]Dpto. de Física Aplicada, Univ. Politècnica de València, Cno. de Vera s/n, 46022 València, Spain

[3]Institute of Applied Physics, Academy of Sciences of Moldova, 2028 Chisinau, Moldova

**Abstract:** The effect of pressure on the optical-absorption edge of $CdIn_2S_4$, $MgIn_2S_4$, and $MnIn_2S_4$ thiospinels at room temperature is investigated up to 20 GPa. The pressure dependence of their band-gaps has been analyzed using the Urbach's rule. We have found that, within the pressure-range of stability of the low-pressure spinel phase, the band-gap of $CdIn_2S_4$ and $MgIn_2S_4$ exhibits a linear blue-shift with pressure, whereas the band-gap of $MnIn_2S_4$ exhibits a pronounced non-linear shift. In addition, an abrupt decrease of the band-gap energies occurs in the three compounds at pressures of 10 GPa, 8.5 GPa, and 7.2 GPa, respectively. Beyond these pressures, the optical-absorption edge red-shifts upon compression for the three studied thiospinels. All these results are discussed in terms of the electronic structure of each compound and their reported structural changes.

---

* Corresponding author, Email: daniel.errandonea@uv.es, Fax: (34) 96 3543146, Tel.: (34) 96 354 4475

**Introduction**

Many AB$_2$X$_4$ compounds with A = Cd, Mg, Mn, Zn; B = Al, Ga, In; and X = O, S are semiconductors that crystallize in the cubic-spinel structure (space group: Fd$\bar{3}$m, No: 227). In this structure the A and B atoms occupy tetrahedral and octahedral sites, respectively [1]. In the last decades, there has been an increasing interest in understanding the high-pressure behavior of ternary compounds belonging to the cubic-spinel family, in particular, since MgAl$_2$O$_4$ became a technologically important compound [2]. Besides the electronic applications of oxide spinels as transparent conducting oxides (TCO) in solar cells and other devices, MgAl$_2$O$_4$ is a prototypical ceramic that plays a significant role in geophysics as a common constituent of the shallow upper mantle [3]. The elastic behavior of its polymorphs can be used to predict seismic velocities, and some of its high-pressure forms have been proposed as constituents of shock-metamorphosed meteorites.

Regarding the sulfide spinel or thiospinel subfamily of cubic-spinels, they are interesting materials for optoelectronic applications given their nonlinear optical properties; e.g. nonlinear optical susceptibility and birefringence [4]. Usually, spinels present some degree of inversion, i.e. the cations can partially interchange their sites in the crystal structure, which makes them defect semiconductors with high concentration of antisite defects. The concentration of these defects can be tuned by means of pressure application, making the thiospinels interesting materials for defect engineering applications [5]. Raman [6] and x-ray diffraction (XRD) experiments [7] as well as pseudo-potential calculations [8, 9] have been performed in order to study the pressure-effects on the crystal structure of AIn$_2$S$_4$ compounds with A = Cd, Mg, Mn. However, no information currently exists on how the electronic band structure of MnIn$_2$S$_4$ and $MgIn_2S_4$ is affected by pressure, existing only limited information for $CdIn_2S_4$ up to 4 GPa [10]. It is the aim of this work to contribute to the understanding of the high-pressure behavior of cubic-spinels, and in particular of the effect of pressure on the electronic and optical properties. In order to address this issue, we have carried out a systematic characterization of the pressure effects on the absorption edge of three different indium thiospinels up to 20 GPa. The obtained results are analysed on the light of previous structural studies in these compounds and its comparison with other isostructural compounds.

**Experimental Details**

High-pressure optical absorption measurements have been performed in spinel-type $CdIn_2S_4$, $MgIn_2S_4$, and $MnIn_2S_4$ at room temperature (RT) up to 20 GPa using a diamond-anvil cell (DAC). The samples used for these measurements were cleaved from single crystals grown by chemical vapor transport using iodine as a transport agent [11]. XRD patterns obtained at ambient conditions confirmed the spinel-type structure of these crystals. For the reported experiments, samples of about 20-30 μm thickness and 100 μm x 100 μm in size were loaded together with a ruby chip in a 250-μm-diameter hole drilled in 45-μm-thick Inconel gasket and inserted between the diamonds of a membrane-type DAC. A methanol-ethanol-water (16:3:1) mixture was used as a pressure transmitting medium and the pressure was determined by the ruby fluorescence technique [12]. The optical-absorption spectra were obtained from the transmittance spectra of the samples which were measured using the sample-in sample-out method in an optical set-up similar to that described in Ref. [13]. At least two independent runs were carried out for each compound in order to check the reproducibility of the obtained results.

**Results and Discussion**

The absorption coefficient ($\alpha$) of the three studied compounds, obtained at several pressures up to 20 GPa, is shown in Figs. 1(a), 2(a), and 3(a). Given the thickness of the studied samples and the stray light level of our spectroscopic system, the highest measurable value of the absorption coefficient is of the order of 5000 cm$^{-1}$, which is a typical value for the low-energy tails of direct absorption edges. The absorption spectra of the three compounds at every pressure exhibit a steep absorption, characteristic of a direct band-gap, plus a low-energy absorption band which overlaps partially with the fundamental absorption. This low-energy absorption band has been observed in related compounds and seems to be related to the presence of defects or impurities [14]. On the other hand, the steep absorption edge exhibits an exponential dependence on the photon energy following the Urbach's law [15]. This dependence is typical of the low-energy tails of direct absorption edges with excitonic effects and has been attributed to the dissociation of excitons in the electric fields of polar phonons or impurities. For values of $\alpha$ below 50 cm$^{-1}$, the photon energy dependence of the absorption coefficient is not exponential but approximately follows a potential law. This has led some authors to propose an indirect character for the fundamental gap of $CdIn_2S_4$ [9]. $MnIn_2S_4$ and $MgIn_2S_4$ are known to have a direct band-gap of 1.95 eV and 2.28 eV [6, 8, 16, 17], respectively. However, according to Meloni *et al.* [9], $CdIn_2S_4$ has an indirect band-gap of 2.2 eV. Our measurements undoubtedly indicate that the three studied compounds are direct band-gap semiconductors, being the determined band-gaps at ambient pressure (10$^{-4}$ GPa) in good agreement with the literature for $MnIn_2S_4$ (1.95 eV) and $MgIn_2S_4$ (2.28 eV). In the case of $CdIn_2S_4$, we have found the band-gap to be 2.35 eV. The differences observed between our work and previous results for $CdIn_2S_4$ could be explained by the fact that previously the presence of the Urbach's tail in the absorption edge was neglected [9]. Therefore, the exponential tail of the direct absorption edge could have been considered as a part of the fundamental absorption, leading to a different characterization of the band-gap and an underestimation of its energy. Recently, Betancourt *et al.* have studied the optical properties of $Cd_{1-x}Mn_xIn_2S_4$ alloys for different samples with x < 0.2 [16], being all the studied samples direct band-gap materials. An extrapolation of the band-gap energies ($E_g$) reported in Ref. [16] to x = 0 (i.e. for $CdIn_2S_4$) yields $E_g$ = 2.45 eV, which confirms our conclusions giving additional support to the hypothesis that $E_g$ was underestimated in Ref. [9].

In Figs. 1(a) and 2(a) it can be seen that the absorption spectra of the spinel phase of $CdIn_2S_4$ and $MgIn_2S_4$ move towards higher energies under compression up to near 10 GPa and 8.5 GPa, respectively. However, in Fig. 3(a) it can be observed that spinel $MnIn_2S_4$ shows an absorption edge that slightly moves towards higher energies only up to near 3 GPa and beyond this pressure it starts to move towards lower energies up to 7.2 GPa. A sudden red-shift jump of the absorption edge is observed in the three compounds at 10 GPa ($CdIn_2S_4$), 8.5 GPa ($MgIn_2S_4$), and 7.2 GPa ($MnIn_2S_4$). This jump in the absorption edge occurs together with a color change of the samples, suggesting the occurrence of a pressure-induced band-gap collapse which could be indicative of a pressure-induced phase transition.

The measured absorption spectra of the three spinels have been analyzed assuming that the band-gap is of direct type and that the absorption edge obeys the Urbach's rule $\alpha = A_0 e^{-(E_g - h\nu)/E_U}$ [15]. In this equation $E_U$ is the Urbach's energy, which is related to the steepness of the absorption tail, and $A_0 = k\sqrt{E_U}$ for a direct band-gap [15], being $k$ a characteristic parameter of each material. In order to simplify the interpretation of our experimental results, we assumed $k$ to be pressure-independent and determined it at the lowest pressure of our experiments. Applying this analysis to our experimental results, we have obtained the pressure dependence for $E_g$ and $E_U$. Figs. 1(b), 2(b) and 3(b) show the obtained results. In $CdIn_2S_4$ and $MgIn_2S_4$, $E_g$ increases linearly upon compression up to 10 GPa and 8.5 GPa, respectively, with a pressure coefficient close to $dE_g/dP = 70$ meV/GPa in good agreement with previous measurements up to 4 GPa in $CdIn_2S_4$ [see empty symbols in Fig. 1(b)] [10]. In contrast, in $MnIn_2S_4$, $E_g$ presents a non-linear evolution with pressure up to 7.2 GPa, showing a maximum around 3 GPa and a pressure coefficient of 20 meV/GPa at ambient pressure. From the analysis of the measured optical absorption spectra, an accused band-gap collapse of around 0.65 eV in $CdIn_2S_4$ at 10 GPa, of around 0.3 eV in $MgIn_2S_4$ at 8.5 GPa, and of around 0.3 eV in $MnIn_2S_4$ at 7.2 GPa has been found. Additional, abrupt changes in the pressure dependence of $E_g$ are observed at 12 GPa, 12 GPa and 9 GPa for $CdIn_2S_4$, $MgIn_2S_4$, and $MnIn_2S_4$, respectively. The above described changes in the band-gap energies and their pressure coefficients for the three compounds suggest either the presence of two phase transitions or that the phase transition taking place in the three compounds is rather slow and there is a coexistence of two phases between 10 GPa and 12 GPa, between 8 GPa and 12 GPa, and between 7.2 GPa and 9 GPa in $CdIn_2S_4$, $MgIn_2S_4$, and $MnIn_2S_4$, respectively.

The occurrence of the described changes correlates quite well with previous high-pressure structural studies [6, 7, 18]. Raman spectroscopy measurements under pressure performed in the three spinels found the onset of a phase transition from the spinel structure to a non-Raman-active phase at 9.3 GPa, 10 GPa, and 7.2 GPa in $CdIn_2S_4$, $MgIn_2S_4$, and $MnIn_2S_4$, respectively [6]. This transition seems to have a martensitic character [19] and is not completed up to 12 GPa in the case of $MgIn_2S_4$. Recently, XRD measurements under pressure have detected the same phase transition and assigned the high-pressure phase to a double NaCl-type structure of the LiTiO$_2$-type [7, 18]. XRD measurements located the onset and completion of the phase transition at slightly different pressures than Raman experiments, being the pressure values for the starting and finishing point of the transition 7.7-12 GPa for CdIn$_2$S$_4$, 8-17 GPa for MgIn$_2$S$_4$, and 6-9 GPa for MnIn$_2$S$_4$ [7, 18]. XRD measurements show that within these pressure ranges, the low- and high-pressure phases coexist. Consequently, the important structural changes taking place during the phase transition could cause the collapse of E$_g$ and the changes in the pressure coefficient of E$_g$ observed in the three spinels, in agreement with what has been previously found in other ternary semiconductors [20].

We would like to comment now the similarities and differences observed in the values of the E$_g$ and its pressure coefficient in the three studied compounds within the pressure range of stability of the spinel phase on the light of previous studies. In the three ternary indium thiospinels, the band-gap energy is very similar to that of the spinel β-In$_2$S$_3$ (around 2.1 eV) [21] which is a defective spinel structure crystallizing above 330ºC with In atoms occupying both tetrahedral and octahedral sites [22]. In fact, β-In$_2$S$_3$ samples show a dark red color similar to that of our ternary thiospinels. X-ray photoelectron spectroscopy measurements evidenced that the topmost valence band in CdIn$_2$S$_4$ was similar to that of β-In$_2$S$_3$ and mainly contributed by S 3$p$ states [23]. The electronic states contributing to the valence band in thiospinels has not been confirmed by theoretical calculations, but are supported by band structure calculations in spinel CdIn$_2$O$_4$ and CdGa$_2$O$_4$ that confirm that the topmost valence band in CdIn$_2$O$_4$ is due to O 2$p$ states [24]. Calculations yield a direct band-gap for CdIn$_2$O$_4$ (CdGa$_2$O$_4$) of around 1.18 eV (1.98 eV). However, the experimentally determined direct gap in CdIn$_2$O$_4$ and MgIn$_2$O$_4$ are similar (around 3 eV [25, 26]) like the direct gaps in our three thiospinels; thus indicating that the energy of the direct gap in AB$_2$X$_4$ spinels is mainly due to the B-X hybridization rather than to the A-X hybridization. This result is reasonable since in the normal spinel each anion X is bonded to three B cations (at octahedral sites) and with only one A cation (at tetrahedral sites); therefore the band structure is mainly formed by the hybridization of the B cation and the X anion.

Band structure calculations in spinel $CdIn_2O_4$ and $CdGa_2O_4$ also show that the lowest conduction band in $CdIn_2O_4$ consists mostly of hybridized In 5$s$ states and Cd 5$s$ states while in $CdGa_2O_4$ it consists mostly of hybridized Ga 4$s$ states and Cd 5$s$ states [24]. This result agrees with those previously obtained by Kawazoe and Ueda, who discussed that the exact value of the direct gap in spinels depends very much on the cation A-cation B interaction because much of the dispersion of the conduction band comes from this interaction [27]. These authors evidenced that metal cations with no $d$ or $f$ states or with filled $d$ and $f$ states would form an extended conduction band. This is the case in spinel-type $In_2S_3$, $CdIn_2S_4$ and $MgIn_2S_4$ so we expect a similar electronic structure with similar band-gaps (around 2.2 eV) and pressure coefficients (around 70 meV/GPa) in the three compounds since the formation of the band-gap is mainly due to the In-S hybridization, as commented before. Consequently, the linear increase of the band-gap is coherent with this interpretation as it reflects the energy shift up of the conduction band (antibonding In 5$s$ character) due to the increase of the bonding-antibonding splitting as the In-distances decrease under pressure. It seems relevant to notice that our result concerning the band-gap pressure coefficient of thiospinels is consistent with the fact that the calculated pressure coefficients of the band-gap of spinel oxides are also very close to each other (around 30 meV/GPa) [24]. The lower absolute value of the pressure coefficient in spinel oxides is most probably related to the lower compressibility of oxides with respect to sulfides.

The importance of the cation A-cation B interaction for the conduction band in $AB_2X_4$ spinels and the nature of the states involved in the lowest conduction band has been confirmed by recent x-ray absorption measurements (XAS) in $\beta$-$In_2S_3$ and by band structure calculations in $CdIn_2O_4$ and $CdGa_2O_4$. XAS measurements have shown that the lowest conduction band in this compound is mainly due to the hybridized In $5s$ orbitals that in $\beta$-$In_2S_3$ are contributed from In atoms at both tetrahedral and octahedral sites of the defect-spinel structure [28]. On the other hand, electronic density of states calculations in the oxide spinels have confirmed that the lowest conduction band is mainly due to Cd $5s$ and In $5s$, and of Cd $5s$ and Ga $4s$ in $CdIn_2O_4$ and $CdGa_2O_4$, respectively [24].

Kawazoe and Ueda also concluded that metal cations having open shell $d$ or $f$ electronic configurations would cause a different conduction band with intrashell transitions in the visible. Therefore we expect, in such cases, smaller band-gaps and smaller pressure coefficients, as we have observed in $MnIn_2S_4$ ($E_g$ = 1.95 eV, $dE_g/dP$ = 20 meV/GPa). This result is reasonable since magnetic-effects interactions could be important in $MnIn_2S_4$ [29] and the $Mn^{3+}$ ions are Jahn-Teller active [30]. Under compression, the spin-state of $Mn^{3+}$ can be modified and the cooperative interactions of $Mn^{3+}$ ions could influence the band structure of $MnIn_2S_4$, which could originate the strongly non-linear behavior observed for $E_g$ in $MnIn_2S_4$. This hypothesis is supported by low-temperature studies, which have shown that in $MnIn_2S_4$ $E_g$ is much less temperature dependent than in the rest of the thiospinels [16]. However, high-pressure band-structure calculations are needed for $MnIn_2S_4$ in order to better understand the non-linear behavior of its band-gap.

Regarding the behavior of $E_g$ in the high-pressure phases, it is important to note that beyond the onset of the phase transition from the spinel phase the band-gap energy has large negative pressure coefficients in the three compounds ($dE_g/dP$ = - 80-110 meV/GPa) up to the completion of the phase transition. However, once the phase transition is finished in all three compounds, the band-gap continues exhibiting a negative linear dependence with pressure, but with a smaller pressure coefficient ($dE_g/dP$ = - 20-30 meV/GPa). On decreasing pressure the spinel phase is recovered in $MgIn_2S_4$ and $MnIn_2S_4$ with almost not detectable hysteresis. This fact is in agreement with the observations made in Raman experiments [6]. However, previous experiments suggested a similar behavior in the Cd thiospinel, a fact that we have not observed in our samples. This irreversibility observed in the optical measurements in $CdIn_2S_4$ can be explained by the creation of defects in the samples under pressure, likely related to the larger ionic radius of Cd with respect to In.

Lets us comment now of the pressure effects on the Urbach's energy. As we mentioned above, this parameter is related with the shape of the absorption edge, so it indirectly gives information on the disorder and the appearance of defects in the spinels. In $CdIn_2S_4$ and $MgIn_2S_4$ at ambient pressure $E_U$ is close to 50 meV, but in $MnIn_2S_4$ it is close to 160 meV, suggesting a higher presence of defects in $MnIn_2S_4$. In $CdIn_2S_4$ and $MgIn_2S_4$, as we increase the pressure, we observed that $E_U$ increases linearly in both compounds up to the onset of the phase transition. Beyond this point, $E_U$ remains practically constant with pressure up to the completion of the phase transition where an additional increase is observed, reaching $E_U$ a value close to 200 meV. This observation agrees with the fact that while samples remain in spinel-type phase the defect concentration increases with pressure, reaching constant values once the samples have transited to a more ordered doubled NaCl-type phase. Against this hypothesis it can be argued that the change of the shape in the absorption edge, that causes the increase of $E_U$, could be the consequence of a pressure-induced direct to indirect band-gap crossover [31]. However, in such a case, the changes of the width of the exponential absorption should be fully reversible, being this not the present case even for samples where pressure was decreased before reaching the onset of the phase transition. Regarding $MnIn_2S_4$, it can be concluded that $E_U$ remains constant upon compression and close to 160 meV for the low-pressure phase. On the other hand, at the phase transition, $E_U$ apparently increase in $MnIn_2S_4$ as in the other two spinels, remaining nearly constant and close to 200 meV in the high-pressure phase. The high $E_U$ value determined for $MnIn_2S_4$ in the low-pressure phase indicates that this compound is a highly disordered spinel even at ambient pressure. We think that the origin of the disorder in $MnIn_2S_4$ is related to the fact that Mn and In have nearly the same ionic radii for the same coordination, which leads to a high degree of inversion (i.e. a large number of Mn and In interchange their positions) in $MnIn_2S_4$ [32], being inversion much less important in $CdIn_2S_4$ and $MgIn_2S_4$ [33].

**Conclusions**

In summary, absorption spectra of $CdIn_2S_4$, $MgIn_2S_4$ and $MnIn_2S_4$ were measured as a function of pressure up to 20 GPa. The analysis of the pressure dependence of these spectra permitted us to obtain the evolution of the band-gaps up to 20 GPa. We have found that within the range of stability of the spinel structure, $CdIn_2S_4$ and $MgIn_2S_4$ behave in a similar way with a linear increase of $E_g$ with pressure, which is in agreement with their expected similar electronic band structures. On the contrary, a non-linear dependence for $E_g$ has been observed in $MnIn_2S_4$. This different behavior could be related with the magnetic character of the Mn cation and its partially filled *d* shells. *Ab initio* band structure calculations on the three thiospinels, in particular for $MnIn_2S_4$, are needed to confirm the nature of the electronic band structure suggested in this work and to better understand the non-linear behavior of the band-gap of $MnIn_2S_4$.

Additionally, abrupt changes were observed in the band-gaps of $CdIn_2S_4$, $MgIn_2S_4$ and $MnIn_2S_4$ at 10 GPa, 8.5 GPa, and 7.2 GPa, respectively. These changes have been interpreted as a consequence of structural changes suffered by the compounds, being correlated with phase transitions previously identified by means of XRD and Raman spectroscopy measurements.

**Acknowledgments:** This study was supported by the Spanish government MCYT under Grants No: MAT2007-65990-C03-01, MAT2006-02279, and MAT2005-07908-C02-01 and the MALTA-Consolider Ingenio 2010 CSD-2007-00045 project. D. Errandonea acknowledges the financial support from the MCYT of Spain and the Universitat de València through the "Ramon y Cajal" program.

**Figure captions**

**Figure 1:** (a) Room temperature optical absorption spectra of CdIn$_2$S$_4$ at different pressures up to 15 GPa. (b) Pressure dependence of band-gap energy (circles) and the Urbach's energy (triangles) in CdIn$_2$S$_4$. Solid (empty) symbols: This work (Ref. 10). The solid vertical lines indicate the onset and completion of the phase transition. I (II) indicates the pressure range of stability of the low- (high-)pressure phase. I + II is the region of phase coexistence.

**Figure 2:** Same as Fig.1 but for MgIn$_2$S$_4$ up to 19.9 GPa. The inset of (a) shows the spectra measured up to 7.3 GPa to avoid the overlap with the spectra measured at higher pressures.

**Figure 3:** Same as Fig.1 but for MnIn$_2$S$_4$ up to 19.5 GPa.

**Figure 1**

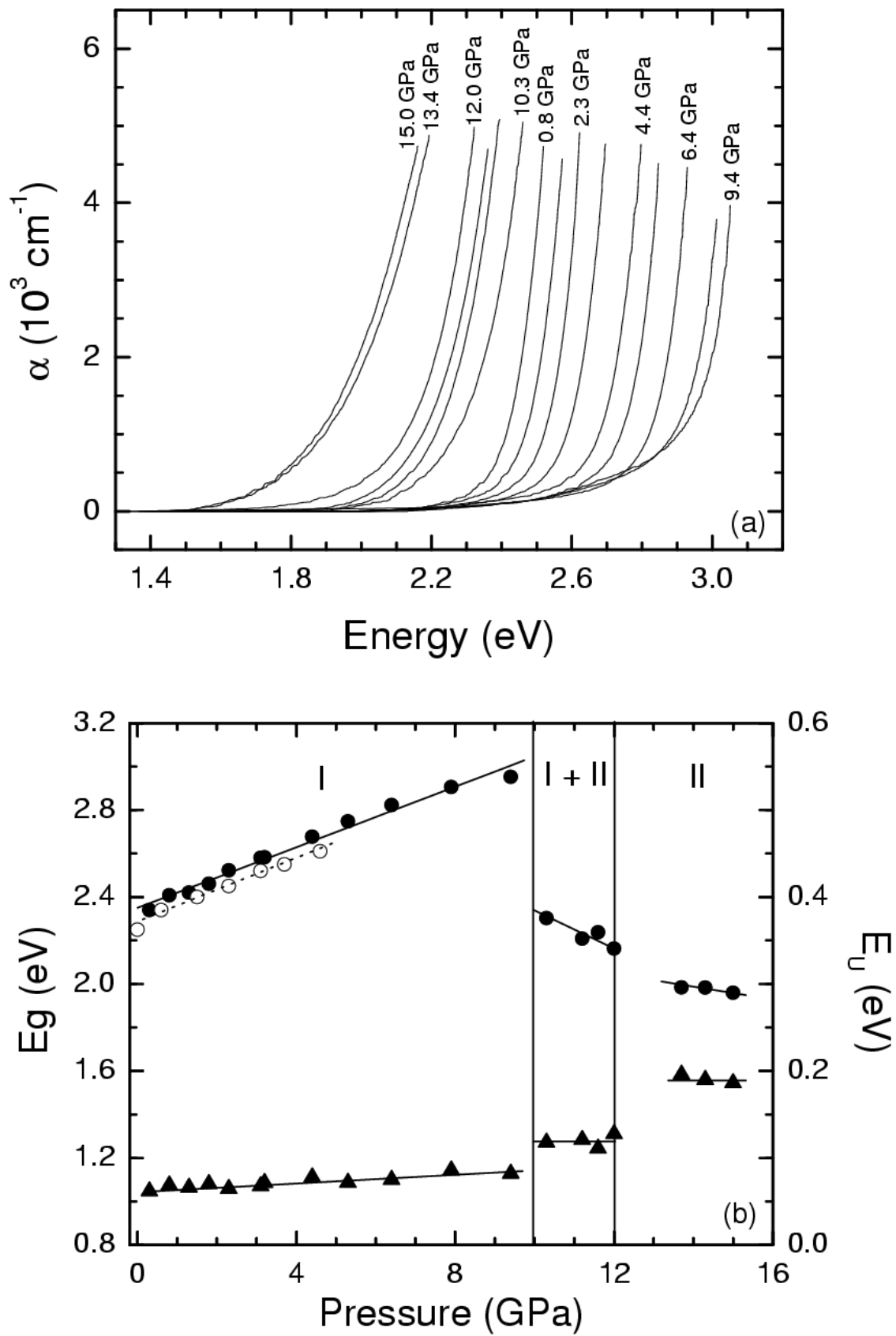

**Figure 2**

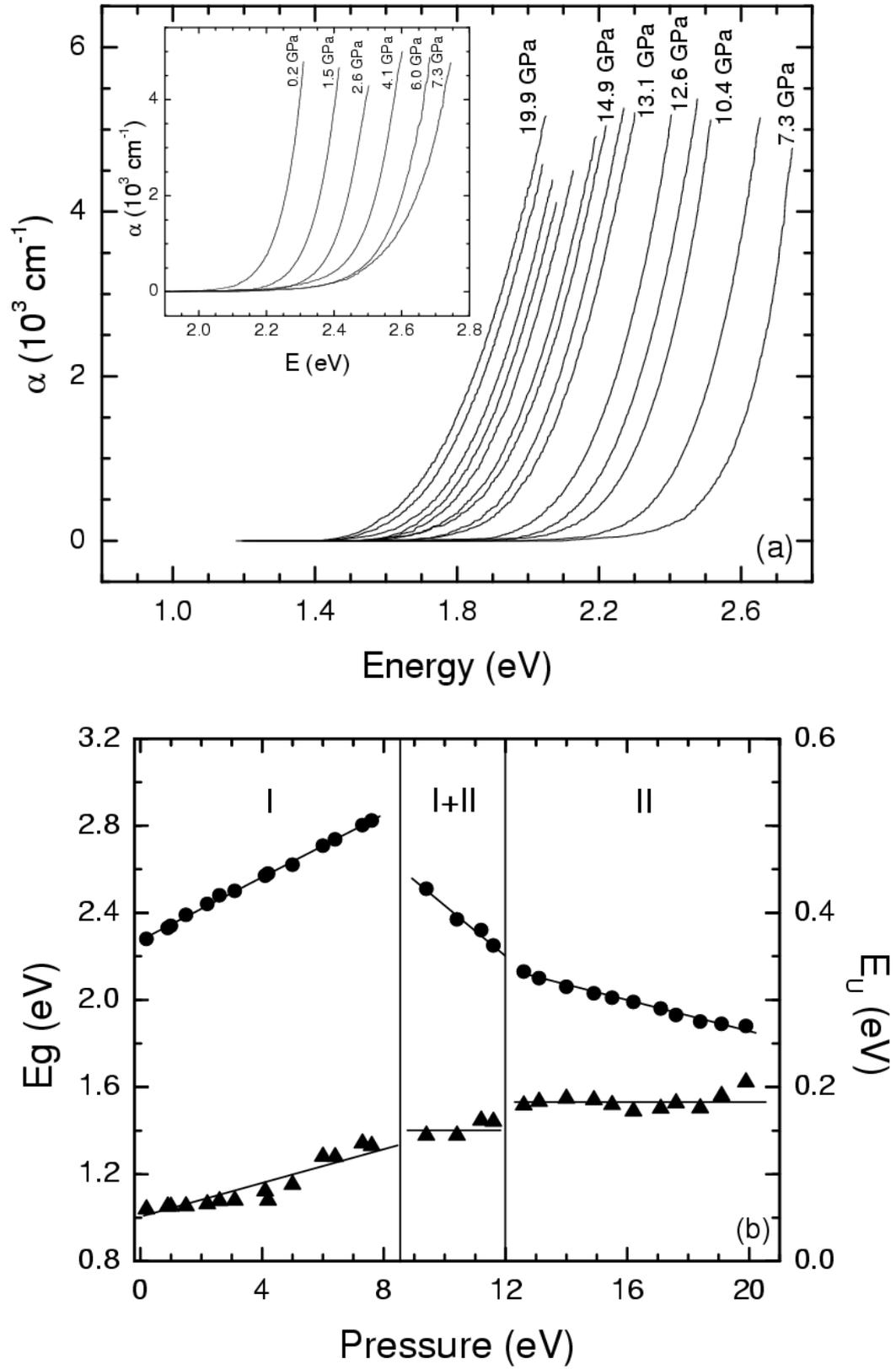

**Figure 3**

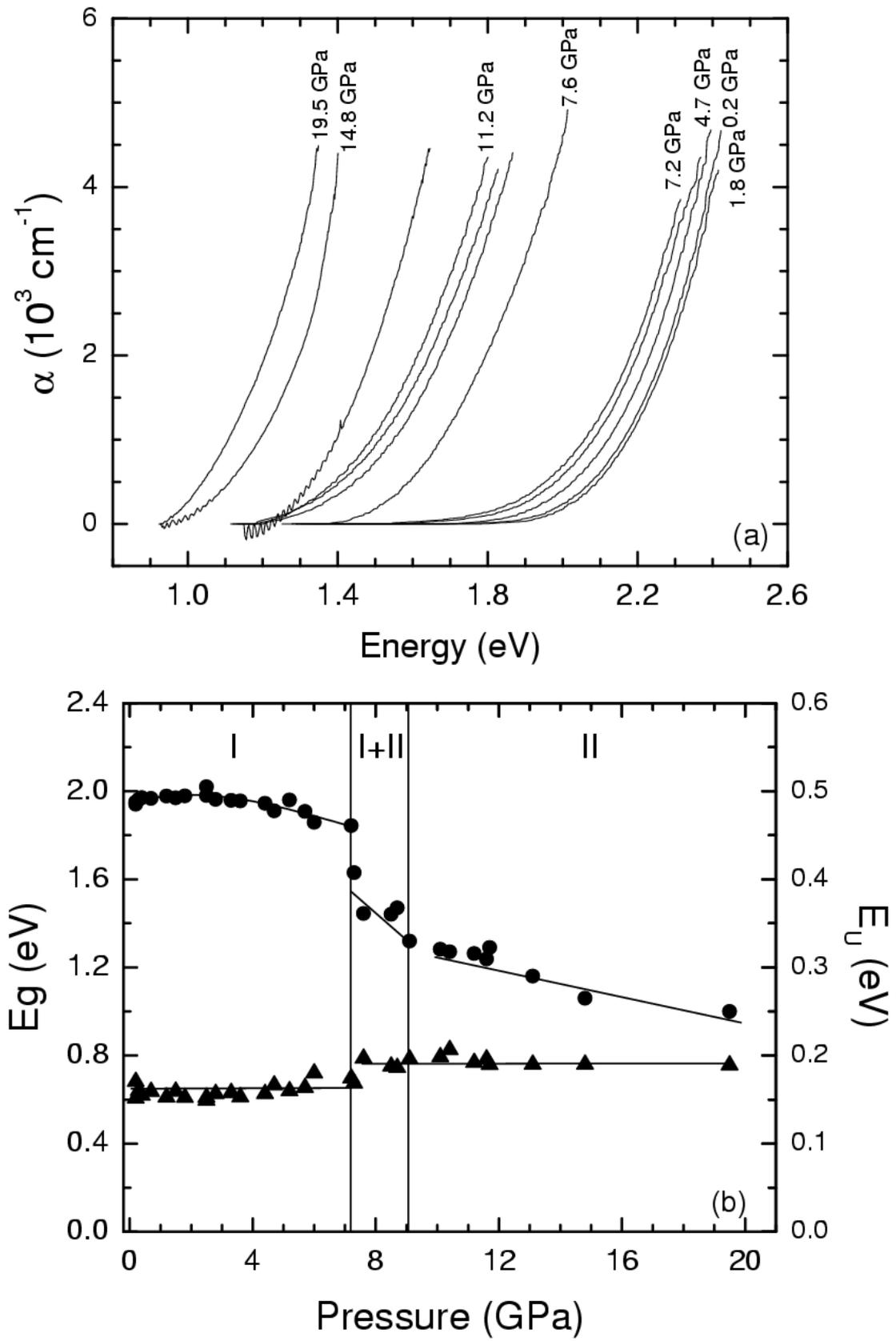